\documentclass[conference]{IEEEtran}
\IEEEoverridecommandlockouts
\usepackage{cite}
\usepackage{amsmath,amssymb,amsfonts}
\usepackage{algorithmic}
\usepackage{graphicx}
\usepackage{textcomp}
\usepackage{xcolor}
\usepackage{braket}
\usepackage{multirow}
\usepackage{caption}
\usepackage{subcaption}
\def\BibTeX{{\rm B\kern-.05em{\sc i\kern-.025em b}\kern-.08em
    T\kern-.1667em\lower.7ex\hbox{E}\kern-.125emX}}
    
\IEEEoverridecommandlockouts

\usepackage{tikz}
\usepackage{hyperref}
\usepackage{lipsum}

\newcommand\copyrighttext{%
  \footnotesize © 2022 IEEE. Personal use of this material is permitted. Permission from IEEE must be obtained for all other uses, in any current or future media, including reprinting/republishing this material for advertising or promotional purposes, creating new collective works, for resale or redistribution to servers or lists, or reuse of any copyrighted component of this work in other works.}
\newcommand\copyrightnotice{%
\begin{tikzpicture}[remember picture,overlay]
\node[anchor=south,yshift=10pt] at (current page.south) {\fbox{\parbox{\dimexpr\textwidth-\fboxsep-\fboxrule\relax}{\copyrighttext}}};
\end{tikzpicture}%
}

\begin{document}

\title{Quantum-classical convolutional neural networks in radiological image classification \\
\thanks{The project/research is supported by the Bavarian Ministry of Economic Affairs, Regional Development and Energy with funds from the Hightech Agenda Bayern.}
}

\author{\IEEEauthorblockN{Andrea Matic}
\IEEEauthorblockA{%\textit{Dependable Perception \& Imaging} \\
\textit{Fraunhofer IKS}\\
Munich, Germany \\
andrea.matic@iks.fraunhofer.de}
\and
\IEEEauthorblockN{Maureen Monnet}
\IEEEauthorblockA{%\textit{Dependable Perception \& Imaging} \\
\textit{Fraunhofer IKS}\\
Munich, Germany \\
maureen.monnet@iks.fraunhofer.de}
\and
\IEEEauthorblockN{Jeanette Miriam Lorenz}
\IEEEauthorblockA{%\textit{Dependable Perception \& Imaging} \\
\textit{Fraunhofer IKS}\\
Munich, Germany \\
jeanette.miriam.lorenz@iks.fraunhofer.de}
\and
\IEEEauthorblockN{Balthasar Schachtner}
\IEEEauthorblockA{%\textit{Department of Radiology} \\
\textit{Department of Radiology}\\
University Hospital, LMU Munich, Germany \\
balthasar.schachtner@med.lmu.de}
\and
\IEEEauthorblockN{Thomas Messerer}
\IEEEauthorblockA{%\textit{Dependable Perception \& Imaging} \\
\textit{Fraunhofer IKS}\\
Munich, Germany \\
thomas.messerer@iks.fraunhofer.de}
}

\maketitle

\copyrightnotice

\begin{abstract}
Quantum machine learning is receiving significant attention currently, but its usefulness in comparison to classical machine learning techniques for practical applications remains unclear. However, there are indications that certain quantum machine learning algorithms might result in improved training capabilities with respect to their classical counterparts - which might be particularly beneficial in situations with little training data available. Such situations naturally arise in medical classification tasks. Within this paper, different hybrid quantum-classical convolutional neural networks (QCCNN) with varying quantum circuit designs and encoding techniques are proposed. They are applied to two- and three-dimensional medical imaging data, e.g. featuring different, potentially malign, lesions in computed tomography scans. The performance of these QCCNNs is already similar to the one of their classical counterparts - therefore encouraging further studies towards the direction of applying these algorithms within medical imaging tasks.
\end{abstract}

\begin{IEEEkeywords}
quantum computing, quantum machine learning, convolutional neural networks, imaging, medical classification, CT scans
\end{IEEEkeywords}

\section{Introduction}

Quantum machine learning (QML) combines two fields currently receiving significant attention: quantum computing and machine learning. The term itself has different meanings -- this paper focuses on applying quantum-enhanced machine learning techniques on classical data. In recent years, work on QML has seen an increased momentum~\cite{Wittek_QML,Schuld_QML,Ganguly_QML}, but the advantage of using QML in comparison to classical machine learning remains unclear~\cite{Schuld:2022lss}. The research has focused on two key directions~\cite{Schuld:2022lss}: to speed-up classical machine learning methods by including quantum-computing-based subroutines or to investigate parametrized, also called variational, quantum circuits~\cite{Cerezo:2020jpv} that can similarly be trained like classical machine learning techniques. As~\cite{Schuld:2022lss} points out, efforts focus very much on `beating' classical machine learning techniques in some aspect, like in a reduced computational complexity or in an increased expressivity of the model.

Most of these studies, however, have been performed on artificial problem settings or on toy datasets, on which also empirical advantages of the QML variants over their classical counterparts were reported. Wide applications to real-world problems or to industrial use-cases are missing. 
More importantly, also a theoretical understanding of cases when quantum algorithms might show an advantage over classical algorithms is lacking. 
This situation partly arises from the lack of error-corrected quantum computers (QC) of large scale (i.e. with many error-corrected qubits), as well as from a missing routine to achieve a Quantum Random Access Memory (QRAM) to read in big amounts of (classical) data. Instead, currently available quantum computers are noisy and therefore referred to as \textit{Noisy Intermediate Scale Quantum} (NISQ) devices~\cite{2018arXiv180100862P}. These NISQ computers feature a limited number of noisy qubits of about 100, show a limited connectivity between the qubits and gate infidelities. This limits the class of quantum circuits and algorithms that could be run on this hardware, restricting them to using a limited number of qubits with a small depth of the circuits. To cope with the limitations of present quantum hardware, the use of hybrid algorithms is most promising, which feature an iteration between quantum computers and classical computing systems. Here, part of the algorithm is run on the quantum computer itself, but other parts still on classical systems. Classical systems could e.g. be used for updating parameter values in variational quantum circuits. Further, it is unlikely that QC will replace classical computers unless for very specific calculations, where the QC will act as quantum processing unit (QPU)~\cite{Humble:2021}. Variational quantum algorithms are therefore natural candidates for quantum algorithms being able to profit from NISQ computers.

In assessing the practical usefulness of these algorithms, it is, however, not yet meaningful to look at real-time speed improvements with respect to classical variants, as the iteration between NISQ computers and classical systems is still too slow in practice. Therefore, within this paper, we concentrate on a different advantage that certain QML algorithms may offer in comparison to their classical counterparts: to result in a more accurate training and generalization in the situation of having small training set sizes available, as proposed in~\cite{Caro:2021mgf}, or by using a smaller number of trainable parameters~\cite{Abbas:2021}.

This situation naturally arises in the context of medical imaging tasks, as in this case imaging data is both available in limited amount and it is difficult or expensive to create more. High accuracy of predictions is desirable, since misclassifications spoil the promises of artificial intelligence (AI) to make clinical decisions faster and more reliable. Many of the rarer diseases would profit from machine-learning methods able to generalize well from small datasets, since the number of patients that can be included into a clinical trial is inherently limited.
Classification tasks on radiological images are conventionally approached with feature-extraction methods such as radiomics \cite{Lambin:2012}  or convolutional neural networks (CNNs)~\cite{Lecun:1998}.
CNNs have achieved superb performance in imaging tasks in general, making them widely used in industrial tasks, but typically require large training datasets. Therefore, we explore if medical classification tasks can profit from the application of hybrid QC-assisted machine-learning methods, particularly in the situation of little training data being present. In this, we build on work by~\cite{Henderson:2020} and~\cite{FOKUS:2021} and apply trainable quantum-classical convolutional neural networks (sometimes also called quanvolutional neural networks) to 2D and 3D medical imaging data featuring different types of lesions or even cancer. 

Our contributions are the following:

\begin{itemize}
    \item We present a hybrid quantum-classical convolutional neural network architecture to identify breast cancer on 2D ultrasound images of the breast and to classify different organs on the axial slices of abdominal computed tomography (CT) scan images.
    \item We extend this method to 3D medical imaging data, demonstrating to our knowledge for the first time the performance of a quantum-classical convolutional neural network on 3D imaging data.
    \item We investigate different quantum encoding schemes to embed the classical data into the quantum circuit, demonstrating that medical datasets behave significantly differently than the simple MNIST~\cite{MNIST} dataset in~\cite{Henderson:2020,FOKUS:2021}.
    \item We present a first study of different designs of the quantum circuit within the quantum-classical convolutional neural network with respect to the training performance.
\end{itemize}

This paper is structured as follows. Section~\ref{related} discusses related work. Section~\ref{background} introduces the background required to understand the architectures of the hybrid quantum-classical convolutional neural networks presented in section~\ref{setup} which also details the setup of our experimental studies. Section~\ref{results} demonstrates the performance of these algorithms on selected medical imaging datasets in simulation.

\section{Related work}

\label{related}
\subsection{CNNs}

CNNs~\cite{Lecun:1998} are a special type of neural networks which are commonly used in computer vision tasks, such as in image classification. A CNN typically consists of a sequence of convolutional and pooling layers, followed by one or more fully connected layers. A convolutional layer identifies image patterns by iteratively convolving the input using $n$ filters with trainable weights. The size of the filter is $k \times k$ for 2D images and $k \times k \times k$ for the convolution of 3D images. The resulting output features are processed by a non-linear activation function.

\subsection{CNNs in radiological imaging}

AI has received widespread attention in the field of medicine and has been shown to reach human performance and to generate unexpected new insights \cite{rajpurkar:2022}.
CNNs are frequently applied to radiological images due to their similarity to conventional computer vision tasks such as classification of photographic images.
Major differences to photographic images are the information acquired by the imaging - in the case of CT the measurement of x-ray absorption in Hounsfield units vs intensity in usually three channels of visible light - and the 3D information acquired by tomographic imaging.
The inherently 3D data has resulted in active research on 3D CNNs for medical imaging \cite{Singh:2020}.
Meanwhile, data curation has been identified as a central bottleneck for the implementation of successful AI in medicine \cite{Wichmann:2020}.
Acquisition of medical imaging data is expensive due to time-consuming expert annotations, imaging costs and privacy concerns. Typical high-quality datasets are therefore comparably small and complex \cite{Prevedello:2019}.
Efficient use of datasets has been identified as an important challenge for the widespread application of AI in medical imaging \cite{Willemink:2020}.

\subsection{Quantum convolutional neural networks}

Given the success of CNNs in image classification tasks, different proposals~\cite{Cong:2019,Kerenidis:2019,Lue:2021,Wei:2022} were developed to apply similar ideas on variational quantum algorithms. In~\cite{Cong:2019} a quantum convolutional neural network (QCNN) with quantum convolutional, quantum pooling and quantum classification layers was proposed, with non-linearities achieved through quantum pooling layers. In comparison to a generic quantum circuit-based classifier, a double exponential reduction of computational complexity was achieved and a more efficient learning demonstrated. In contrast, \cite{Li:2020} suggested a QCNN consisting of sequences of quantum convolutional and quantum classification layers without any pooling layers. Good classification performance is reached on the MNIST~\cite{MNIST} and GTSRB~\cite{GTSRB} datasets, but the method requires QRAM, which is not yet available on existing QC.
On NISQ devices, the data input sizes are challenging. Therefore, \cite{Park:2021} restricts to using only one- and two-qubit gates and to reduce the dimension of the input data by applying either bilinear interpolation techniques, a principal component analysis or an autoencoder.

\subsection{Hybrid quantum-classical convolutional neural networks}

Alternatively to mapping the full idea of a CNN to a quantum variant, \cite{Henderson:2020,FOKUS:2021,Liu:2019} explore to implement only some of the convolutional layers as quantum convolutional layer, while keeping the remaining parts of the architecture as classical layers, thus obtaining a truly hybrid quantum-classical algorithm. The advantage of this approach is that typically no QRAM is required. The proposed architectures differ in whether the quantum convolutional layer is trainable or not. Specifically, \cite{Henderson:2020} proposes a quanvolutional layer which consists of untrainable, randomly selected gates. In comparison to a conventional CNN, this architecture shows an improved performance, but does not outperform a classical variant with a classical random layer added in front. As encoding scheme, \cite{Henderson:2020} uses threshold encoding as described in section~\ref{sec:qcnn_background}, which might not be well suited for gray-scale or more complicated imaging data.

Alternatively, \cite{FOKUS:2021} explores different encoding schemes in addition to the threshold encoding, and extends the untrainable quanvolutional layer to a trainable quantum convolutional layer, although the rest of the CNN remains again classical. It is demonstrated that other encoding schemes such as the flexible representation of quantum images (FRQI)~\cite{Le:2011} and novel enhanced quantum image representation of digital images (NEQR)~\cite{Zhang:2013} encoding might be better suited than threshold encoding in certain cases, where the authors explored the MNIST dataset. In contrast to~\cite{Henderson:2020}, their quantum convolutional layer returns a vector of measured qubit values in the Z-basis, whereas the output of the quanvolutional layer of~\cite{Henderson:2020} is a sum over the qubits being measured as $\ket{1}$ in the state vector. Both~\cite{Henderson:2020} and \cite{FOKUS:2021} explore the performance of their networks in simulation only due to unfeasibly long running times on quantum hardware. Furthermore, \cite{FOKUS:2021} only processes a small dataset size of 200 images to reduce computational time in simulation as well. Due to this, also no hyperparameter tuning was performed. 

\section{Background}

\label{background}

\subsection{Details of the CNNs within this work}

CNNs used within this work feature a simple architecture of a sequence of convolutional layers (just one in case of 2D data), followed by activation functions.
For this, the ReLU function is used. With a consecutive pooling layer the size of the feature maps is reduced and overfitting can be avoided. Fully connected layers are used to perform the classification. The number of outputs of the final layer corresponds to the number of different classes used in the training. The predicted class is determined by the class with the largest output score.

\subsection{Details of the hybrid QCCNNs}
\label{sec:qcnn_background}

The hybrid QCCNNs used within this work apply quantum convolutional layers in addition or instead to the conventional convolutional, pooling and fully connected layers of a CNN. These consist of one or more variational quantum circuits acting as convolutional filters. In the following, the different components of a variational quantum circuit are described. 
\newline

\subsubsection{Encoding}

The encoding feature map, in the following referred to as encoding, describes how different (data) input values are mapped onto the qubits of a quantum circuit. Different techniques exist for this purpose. The encoding technique determines how many qubits are required in the quantum circuit. This work only uses techniques in which one input value is encoded onto exactly one qubit. Therefore, the number of required qubits is equal to the number of inputs processed in one convolution, which is given by the filter size of $k \times k$ for 2D images and $k \times k \times k$ for 3D images. The following encoding techniques are considered in this work:

\begin{itemize}
\item \textit{Threshold encoding}:

This encoding technique was introduced in~\cite{Henderson:2020}. Using a certain threshold $t$, an input $x$ is encoded to the quantum state $\ket{0}$ if $x < t$, and otherwise to the state $\ket{1}$. 

\item \textit{Angle encoding}:

In the angle encoding~\cite{Stoudenmire:2016}, rotation gates with angles based on the input values $x$ are used for encoding. Similarly to \cite{Schuld_superv}, which encodes the input through a rotation around the $Y$-axis by the angle $x$, we perform the encoding using $X$-axis rotations. In our experiments, the qubits are initially in state $\ket{0}$. This results in the quantum state $R_X (x) \ket{0}$ after encoding.

\item \textit{Higher order encoding}:

This encoding technique was proposed in \cite{Havlicek:2018} and uses two-qubit gates in addition to single-qubit gates. This technique leads to an entangled encoding feature map. The qubits are initially in state $\ket{0}$ and are transformed using a Hadamard gate and a $Z$-axis rotation $R_Z(x_n)$, where $x_n$ denotes the $n$-th input. Afterwards, an entangling operation $R_{ZZ}(\phi_{ij})$ is applied to every qubit pair $i$ and $j$. This operation consists of a CNOT gate, a rotation $R_Z(\phi_{ij})$, and another CNOT gate. The rotation $R_Z(\phi_{ij})$ is applied on the $j$-th qubit and we use $\phi_{ij} = x_i*x_j$. Following the naming convention of \cite{Abbas:2021}, we refer to this technique as \textit{higher order encoding}. Although this encoding is more complicated than the other two methods, it is particularly interesting for our studies: in \cite{Havlicek:2018} it was shown that for support vector machines quantum advantage can only be achieved if the encoding feature map is difficult to simulate classically. Therefore, in our experiments we study the performance of this encoding technique for QCCNNs. 
\end{itemize}

\subsubsection{Circuit design}
The circuit design describes which combination of rotation and entangling gates is applied to the qubits after the encoding. Similarly to the motivation for the higher order encoding, we choose circuits with entangling gates across all adjacent qubits. Thereby, we can make use of superposition and interference effects in the circuit. Applied to imaging data, this may enable us to find more complex features than with classical CNNs. 

In a trainable quantum circuit, the rotation gates, or even the whole circuit design, are optimized during the training. In our experiments the circuit design is fixed and only the rotation angles are optimized. The angles are initialized randomly and are then iteratively updated using a classical optimizer.

We study two different circuit designs: the \textit{basic entangling} and the \textit{strongly entangling layer}. In the basic entangling layer each qubit $i$ is rotated by a trainable angle $\theta_i$ around one certain rotation axis. For this, we use $X$-axis rotations. After that, a sequence of entangling gates is applied. The strongly entangling layer applies single-qubit rotations around all three axes $X$, $Y$ and $Z$ before a sequence of entangling gates. Therefore, it contains three times more trainable parameters than the basic entangling layer. We use CNOT gates for entanglement in both layer types.

\subsubsection{Measurement and output features}
At the end of the quantum convolutional layer, qubit measurements are performed and the results are stored in an output feature map. We use the same measurement strategy as in \cite{FOKUS:2021}: The results are measured in the $Z$-basis and the number of feature maps corresponds to the total number of qubits in the quantum convolutional layer. The expectation value from measuring each qubit is stored directly in the output feature map - in contrast to a classical convolution, in which a summation over the convoluted patch is performed. The outputs are not transformed with an activation function in the end.

\section{Experimental setup}

\label{setup}

To achieve a fair comparison between classical ML techniques and QML, \cite{Potempa:2021} argues that quantum-enhanced networks beat classical networks if built with a similar number of parameters. Our hybrid QCCNNs and CNNs are thus designed to have a total number of parameters in the same order of magnitude. For this purpose, we select and replace one convolutional layer from the CNN by a variational quantum circuit to create a QCCNN. The number of input and output features in the classical and quantum convolutional layers is designed to be the same. This results in networks with a similar number of parameters in the classical and quantum convolutional layers, and exactly the same number of parameters in the rest of the networks. In the quantum convolutional layer, the number of parameters is determined by the encoding scheme and the quantum circuit design. We apply our QCCNNs to two 2D medical imaging datasets, including breast ultrasounds and abdominal CT scans, and to one 3D medical dataset consisting of lung CT scans. These datasets and the corresponding architecture setups are described in the following.

\subsection{Datasets}
\subsubsection{MedMNIST Datasets (2D datasets)}

\paragraph{Breast ultrasound images (BreastMNIST)}
This dataset is part of the MedMNIST datasets~\cite{Yang:2021}. It was pre-processed from~\cite{Al-Dhabyani:2020} consisting of breast ultrasound images from 600 female patients with an average image size of $500 \times 500$ pixels. The images show normal, benign and malignant lesions.
In MedMNIST, the images were down-scaled to a low-resolution of $28 \times 28$ and categorized into non-malignant
and malignant classes by merging the normal and benign states. It is a relatively small dataset based on 546 training images and 78 validation images. The end task is a binary classification of breast cancer. This dataset is well suited for our study, because the low resolution of the images allows us to avoid too long processing times.
The small number of training data points serves as basis for assessing the performance of QCCNNs in the presence of little data. The data is normalized to a mean of 0 with a standard deviation of 1.

\paragraph{Abdominal CT scans (OrganAMNIST)}
This dataset equally belongs to the MedMNIST benchmark~\cite{Yang:2021}. The images are based on 3D abdomen CT scans from the Liver Tumor Segmentation Benchmark (LiTS)~\cite{Bilic:2019}, which gathers images of different pre- and post-therapy liver tumor diseases. The original images are diverse in resolution and image quality. They were cropped into 2D slices (in axial views) and resized to $1 \times 28 \times 28$ in the MedMNIST dataset. The task is to perform multi-class classification of 11 organs.
The organ labels in MedMNIST were obtained from the bounding box annotations of another study~\cite{Xu:2019}. For our study, we take a subset of 1,000 and 600 images from the training and validation sets, respectively. The data is normalized to a mean of 0 with a standard deviation of 1.

\subsubsection{Lung computed tomography scans (lung-nodule dataset, 3D)}

\begin{figure*}[t!]
\centering

\begin{subfigure}{0.4\textwidth}
    \centering
    \includegraphics[width=\textwidth]{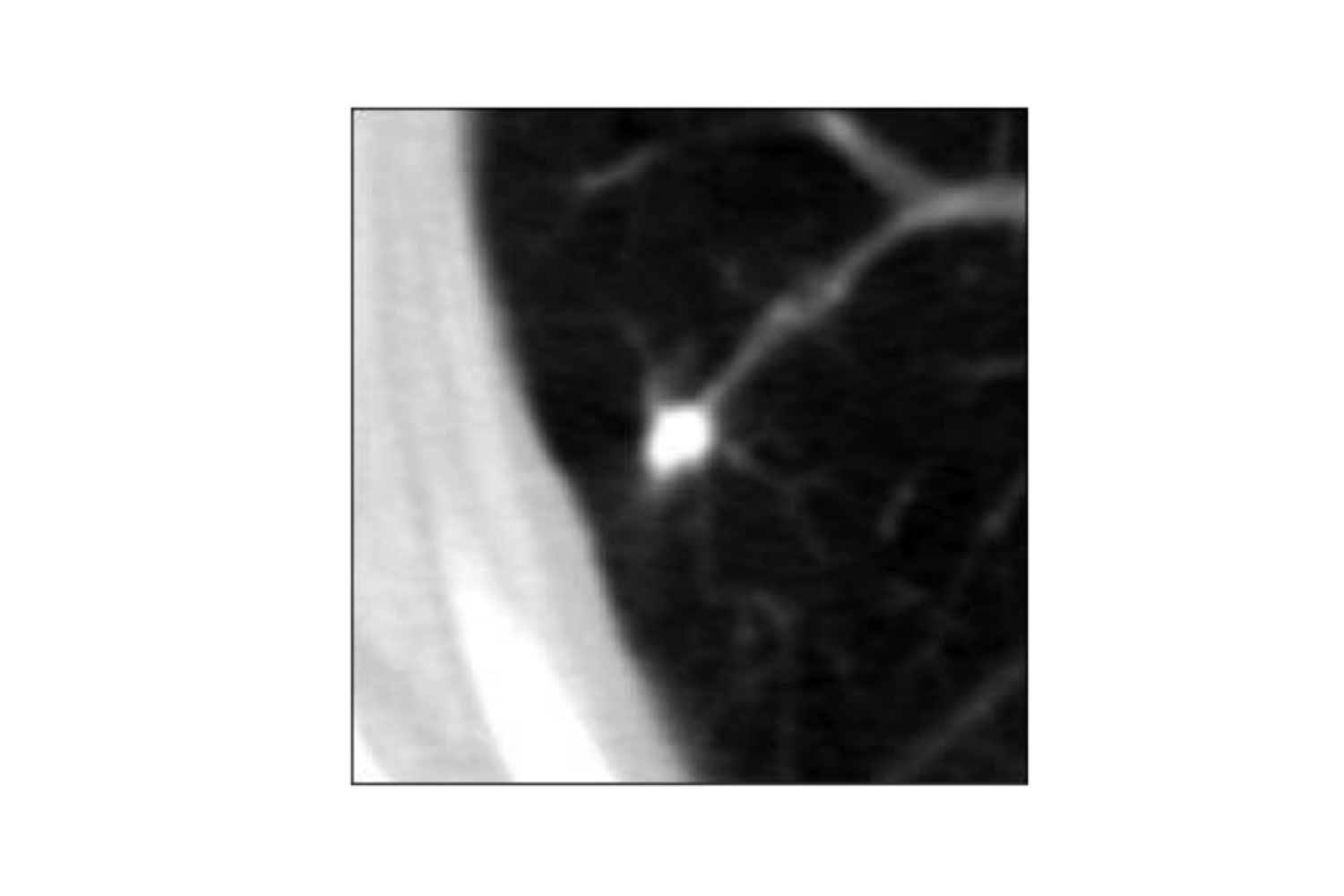}
    \caption{Benign lung nodule}
\end{subfigure}
\qquad
\begin{subfigure}{0.4\textwidth}
    \centering
    \includegraphics[width=\textwidth]{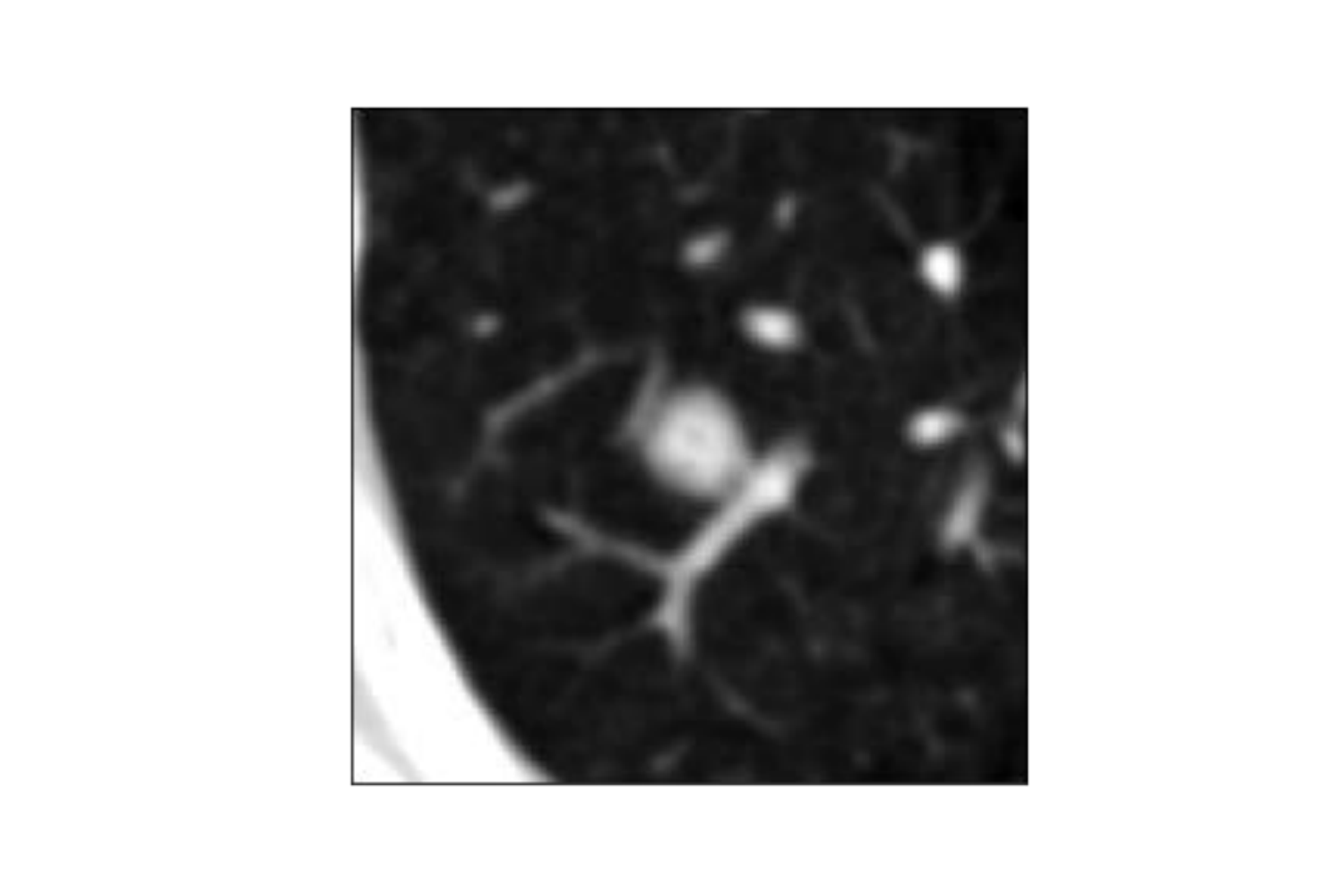}
    \caption{Malignant lung nodule}
\end{subfigure}

\caption{\label{fig:example_3D}Example images for the 3D lung-nodule dataset. Each example shows the central slice in axial direction of one the $6\times6\times6$ cm$^3$ cubes extracted from the LIDC-IDRI dataset used for training and testing the 3D architectures.}

\end{figure*}

We derive a 3D dataset from CT images of the LIDC-IDRI\cite{LIDC:2011} dataset\cite{LIDC_data:2015} provided by the cancer imaging archive\cite{tcia:2013}.
The dataset consists of CT scans of the lung from examinations of lung cancer screening. It contains lesions annotated by multiple expert radiologists.
For the selection of lesions, we consider lesions with a consensus of at least 50\% of the radiologists.
This way for 2630 lesions a bounding box of the localization and assessments of dignity was obtained.
For each lesion a volume of $6\times6\times6$  cm$^3$ was extracted around the center of the bounding box and resampled to a common resolution of $128\times128\times64$ voxels using cubic splines.
The label for the binary classification task is obtained by requiring the median estimate of dignity (scale from 1-5) of all radiologists to be larger than 3.
Example images of the derived dataset are shown in \figurename~\ref{fig:example_3D}.

\subsection{Architecture}

We define two network architectures in this study: one for the two 2D datasets, BreastMNIST and OrganAMNIST, and a second one for the 3D lung-nodule dataset. The two architectures differ in the size of the networks and in the position of the quantum convolutional layer within the network.
\newline

\subsubsection{2D datasets}

\begin{figure*}[t!]
    \centering
    \includegraphics[width=0.7\textwidth]{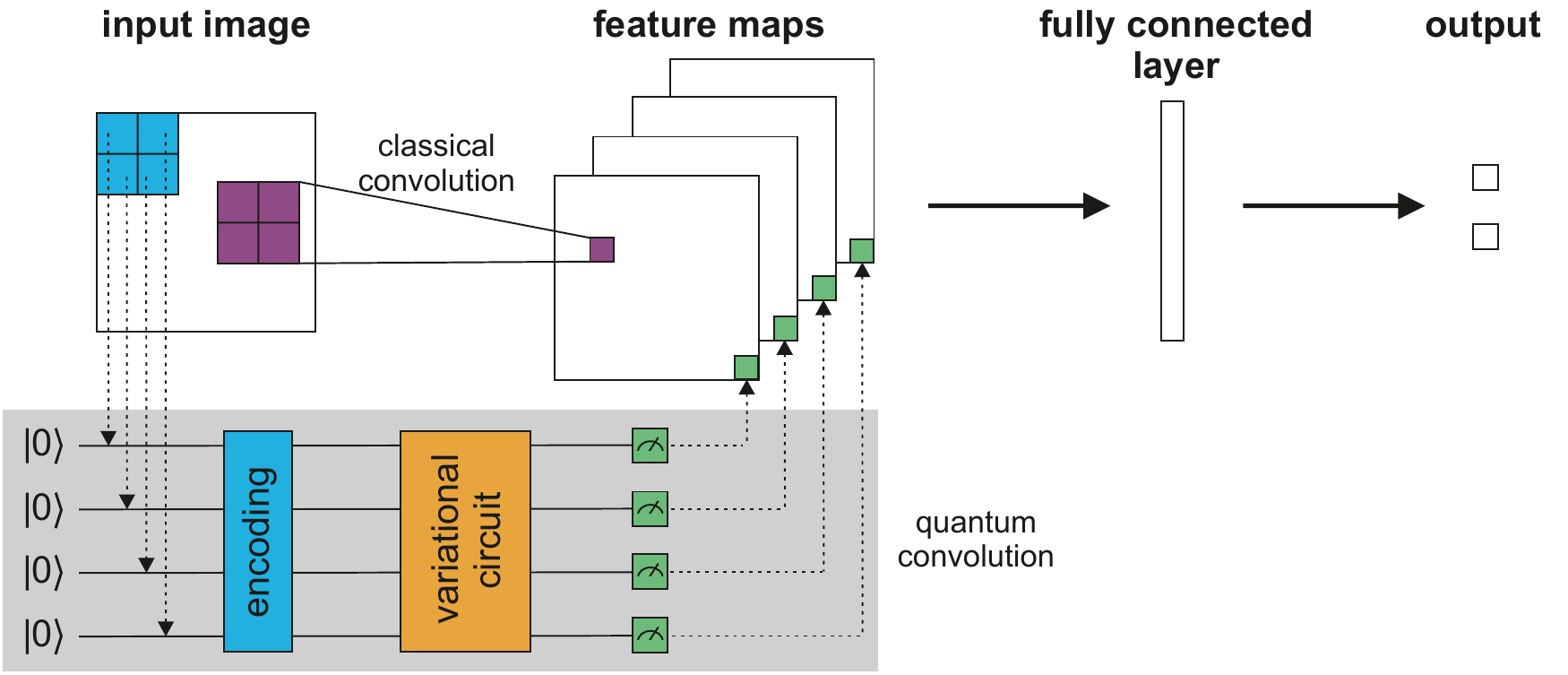}
    \caption{Architecture sketch of the classical and hybrid quantum-classical model used for the 2D datasets.}
    \label{fig:architecture_2D}
\end{figure*}

The architecture used for the 2D datasets is shown in \figurename~\ref{fig:architecture_2D}. The classical network simply consists of one classical convolutional layer with filters of size of $2 \times 2$ and a stride of 2, directly followed by a fully connected layer. Its hybrid quantum-classical twin consists of a quantum convolutional layer with the same filter size and stride, followed by a fully connected layer. The quantum convolutional layer consists of one quantum circuit, i.e. just one filter. With the chosen filter size and encoding scheme, the number of required qubits in the circuit is 4, thus producing 4 output feature maps. To have the same number of output features in the classical CNN, we apply 4 filters to convolve the input image.

\begin{figure*}[h!]
    \centering
    \includegraphics[width=0.8\textwidth]{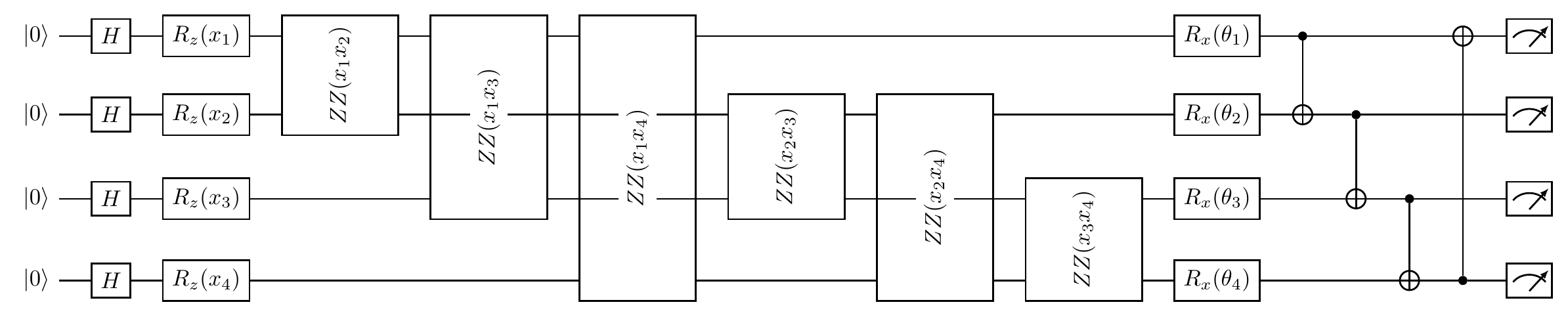}
    \caption{One of the circuits tested for the 2D datasets: The input image is transformed with the higher order encoding, followed by a basic entangling layer. The first four $R_Z$ gates behind the Hadamard gate represent rotations around the $Z$-axis with rotation angles $x_i$, where $x_i$ denotes the $i$-th input. At the other $R_{ZZ}$ gates the rotation angles $x_i x_j$ are applied. }
    \label{fig:circuit_2D}
\end{figure*}

We test three different QCCNN designs for these datasets using the following encodings and architectures:

    \begin{itemize}
        \item Higher order encoding and a basic entangling layer. The corresponding circuit is shown in \figurename~\ref{fig:circuit_2D}.
        \item Higher order encoding and a strongly entangling layer.
        \item Threshold encoding and a strongly entangling layer. Since the datasets are normalized to a mean of 0, we choose as threshold $t = 0$.
    \end{itemize}

Each rotation gate in the basic entangling layer contains one rotation angle $\theta_X$, while the strongly entangling layer contains three rotation angles $\theta_X$, $\theta_Y$ and $\theta_Z$. Thus, we have 4 and 12 trainable parameters for the experiments with the basic and strongly entangling layers, respectively. In comparison, the corresponding classical convolutional layer has 20 trainable parameters. The number of parameters in the latter is given by \((k^2 \times c + 1) \times n\), with $k^2$ being the filter size, $c$ the number of channels in the input image and $n$ the number of filters. The parameter counts in the classical and the different quantum convolutional layers are summarized in Table~\ref{tab1}. Apart from these, other 8,635 trainable parameters in the networks are obtained from the fully connected layer.
\newline

\begin{table}[htbp]
\caption{2D Convolutional Layer Parameter Count}
\begin{center}
\resizebox{0.5\textwidth}{!}{
\begin{tabular}{|c|c|c|c|}
\hline
\textbf{}&\multicolumn{3}{|c|}{\textbf{Convolutional layer type}} \\
\cline{2-4} 
\textbf{}& \textbf{\textit{Classical}}& \textbf{\textit{Basic entangling}}& \textbf{\textit{Strongly entangling}} \\
\hline
\multirow{2}{*}{\# parameters}& \((2^2\cdot 1 + 1) \cdot 4\) 
 & \multirow{2}{*}{Total: 4} & \multirow{2}{*}{Total: 12} \\
& Total: 20 & & \\

\hline
%\multicolumn{4}{1}{Number of trainable parameters in the classical and quantum convolutional layers. Encoding makes no difference. Describe calculation.}

\end{tabular}
}
\label{tab1}
\end{center}

\end{table}

%\onecolumngrid

%\twocolumngrid

\subsubsection{3D dataset}

\begin{figure*}[t!]
    \centering
    \includegraphics[width=0.9\textwidth]{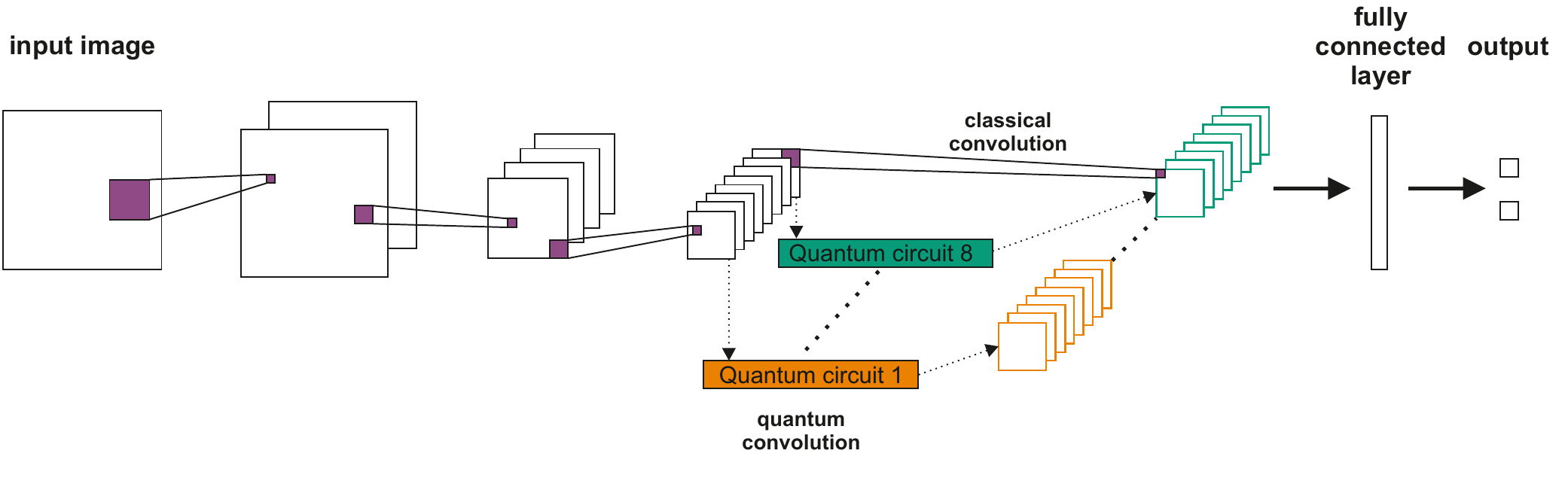}
    \caption{Architecture sketch of the classical and hybrid quantum-classical model used for the 3D dataset.}
    \label{fig:architecture_3D}
\end{figure*}

The 3D task is more complex and therefore requires a larger architecture, which is shown in \figurename~\ref{fig:architecture_3D}. The classical network consists of four 3D convolutional layers followed by a fully connected layer. For its quantum analog, we chose to replace the final convolutional layer in order to work with a reduced number of input features in the quantum circuit. This choice is due to the otherwise prohibitively long processing times of the hybrid algorithm. Consequently, the inputs to the quantum convolutional layer are not the normalized image data, but instead the output values obtained after the third classical convolutional layer processed by the ReLU activation function. 

For the first classical convolutional layer, we use 2 filters with a filter size of $5 \times 5 \times 5$ and a stride of 2. In the second and third layer we use 4 and 8 filters, respectively, with a filter size of $2 \times 2 \times 2$ and a stride of 1. In all three layers, all inputs are convolved by all filters. After each of these convolution layers, we perform batch normalization and apply a pooling layer with a filter size and stride of 2 as well as a dropout of 0.2. The latter is used to prevent overfitting. After the fourth convolutional layer we apply batch normalization and a dropout of 0.5. 

Since the third convolutional layer results in 8 feature maps, the quantum convolutional layer consists of 8 separate quantum circuits, where each of them processes exactly one feature map. The circuits have the same design and differ between each other only by the initial weight initializations. As we work here with 3D convolutions, the required number of qubits for each circuit is $2^3 = 8$. With 8 qubits and 8 feature maps, the quantum convolutional layer leads to 64 output feature maps in total. 

To ensure the same number of output features and a similar number of trainable parameters between the classical and the hybrid quantum-classical setups, we use 64 filters in the fourth classical layer. These filters are divided into 8 groups, such that each input feature map is convolved by one group of filters.

In contrast to the QCCNNs used for the 2D datasets, we use here angle encoding instead of threshold or higher order encoding: 
As we want to keep as much detailed information as possible after encoding, we chose not to use threshold encoding, which maps the inputs only onto the two states $\ket{0}$ and $\ket{1}$. 
On the other hand, in the angle and higher order encoding the input values are encoded through corresponding rotation angles, therefore leading to more information being passed to the quantum circuit. Due to the additional entangling properties, the higher order encoding is in general a very promising encoding technique. However, in first preliminary studies we observed a poor performance for this particular dataset and network architecture. The reasons for this need to be investigated in future research. Therefore, we decided to use angle encoding instead.

\begin{figure*}[tbh!]
    \centering
    \includegraphics[width=0.7\textwidth]{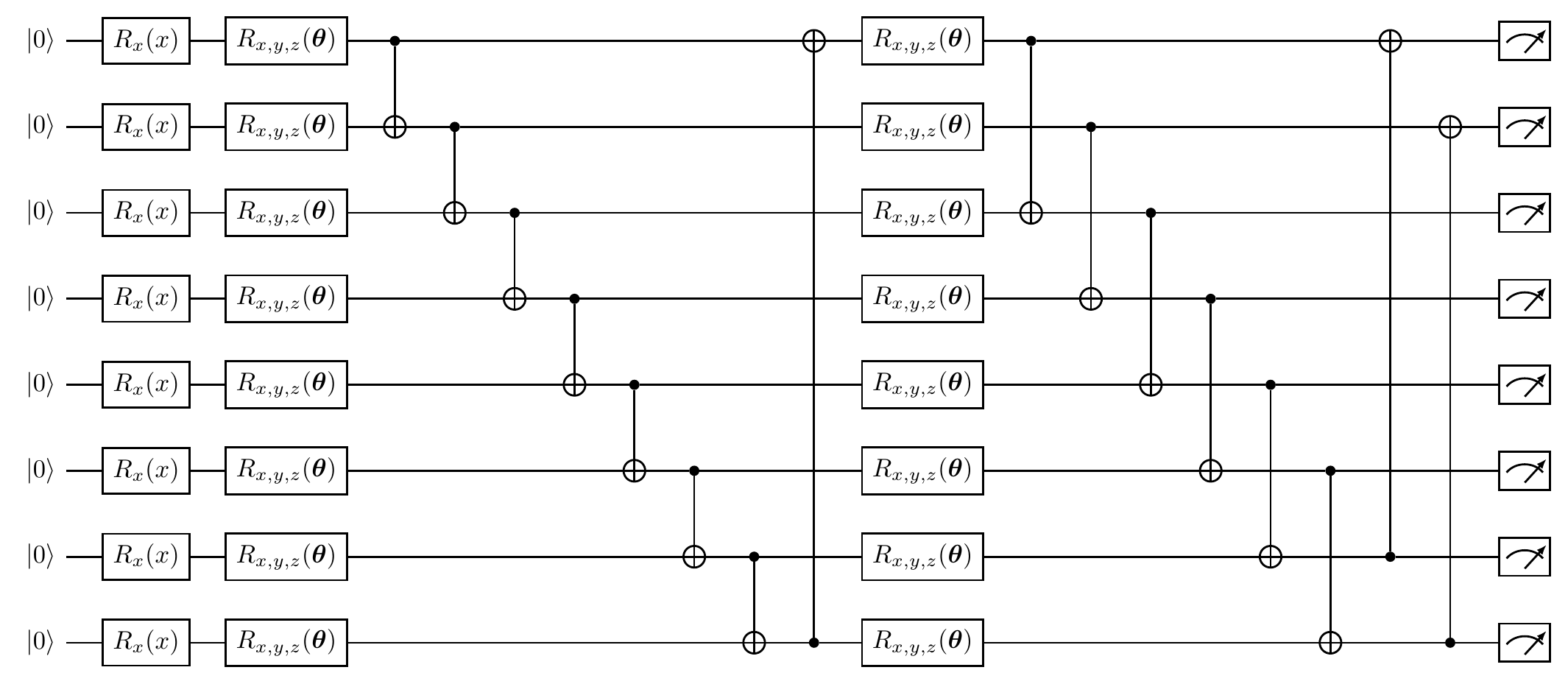}
    \caption{One of the circuits tested for the 3D datasets: The input image is transformed with angle encoding, followed by two strongly entangling layers. In the other  with one strongly entangling layer the circuit consists of the angle encoding, followed only by the first sequence of rotation and CNOT gates.}
    \label{fig:circuit_3D}
\end{figure*}

In this work we compare the performance of the CNN with two QCCNNs:

    \begin{itemize}
        \item Angle encoding and 1 strongly entangling layer.
        \item Angle encoding and 2 strongly entangling layers. The corresponding circuit is shown in \figurename~\ref{fig:circuit_3D}.
    \end{itemize}

Thereby we want to study whether adding a second entangling layer can increase the performance compared to the setup with only one entangling layer. With 8 qubits one obtains 24 and 48 trainable parameters per circuit for the 1-layer and 2-layer setups, respectively. As we use 8 circuits as filters, the total number of parameters is 192 and 384, respectively. For comparison, the corresponding convolutional layer in the classical setup has 576 trainable parameters since in the 3D case, the number of parameters in the convolutional layer is given by \((\frac{k^3 \times c}{g} + 1) \times n\), with $k^3$ being the filter size, $c$ the number of channels in the input image, $g$ the number of grouped filters and $n$ the number of total filters. The trainable parameters in the quantum convolutional layer and in the corresponding convolutional layer of the classical CNN are summarized in Table~\ref{tab2}. In all setups, there are additional 1,894 trainable parameters coming from the fully connected layer and the other three convolutional layers.

\begin{table}[htbp]
\caption{3D Convolutional Layer Parameter Count}
\begin{center}
\resizebox{0.5\textwidth}{!}{
\begin{tabular}{|c|c|c|c|}
\hline
\textbf{}&\multicolumn{3}{|c|}{\textbf{Convolutional layer type}} \\
\cline{2-4} 
& \multirow{2}{*}{\textbf{\textit{Classical}}}& \textbf{\textit{Strongly ent.}}& \textbf{\textit{Strongly ent.}} \\
&  & 1 layer & 2 layers \\

\hline
\multirow{2}{*}{\# parameters}& \((\frac{2^3 \cdot 8}{8} + 1) \cdot 64\)& $24 \cdot 8 $&  $48 \cdot 8 $\\
& Total: 576 & Total: 192 & Total: 384 \\

\hline
%\multicolumn{4}{1}{Number of trainable parameters in the classical and quantum convolutional layers. Encoding makes no difference. Describe calculation.}

\end{tabular}
}
\label{tab2}
\end{center}

\end{table}

%\onecolumngrid

%\twocolumngrid

\section{Results}

\label{results}

Pennylane~\cite{Pennylane:2018} and PyTorch~\cite{Pytorch:2019} are used for conducting the experiments. Pennylane is a python library for building and running quantum machine learning algorithms. For all experiments, we use the \texttt{default.qubit} device of Pennylane, which performs a simple simulation of qubit-based quantum circuits. We took the measurement results from the analytically calculated state vectors instead of calculating the expectation values stochastically with a certain number of shots. Furthermore, noise effects are neglected in the simulations.   

As performance measure, we focus on the evolution of the cross-entropy loss and the accuracy on the training and validation datasets.
We plot the averaged metric values by epoch over a total of 5 networks initialized with different seeds for the angles of the rotation gates and the classical weights. The error band corresponds to the average metric plus or minus its standard deviation. We train all networks for 20 epochs using the Adam optimizer with a learning rate of 0.001 and a batch size of 8, 16 and 64 for the 2D BreastMNIST, the 2D OrganAMNIST and the 3D lung-nodule datasets. We do not do any extensive hyperparameter tuning in view of the resource-intensive experiments.
\newline

\subsubsection{2D datasets}

%\onecolumngrid

\begin{figure*}[h!]
\centering

\begin{subfigure}{0.4\textwidth}
    \centering
    \includegraphics[width=\textwidth]{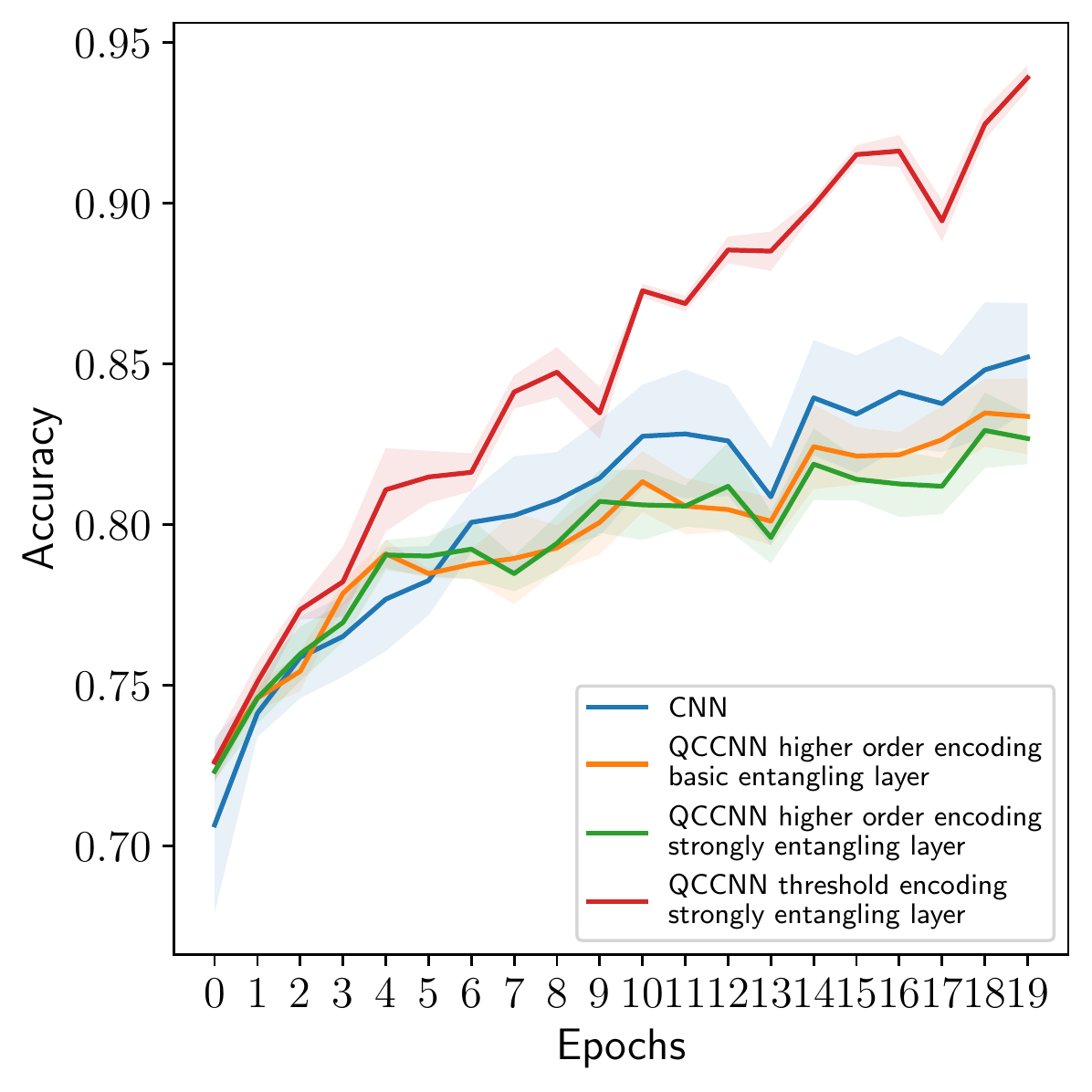}
    \caption{Training accuracy}
\end{subfigure}
\qquad
\begin{subfigure}{0.4\textwidth}
    \centering
    \includegraphics[width=\textwidth]{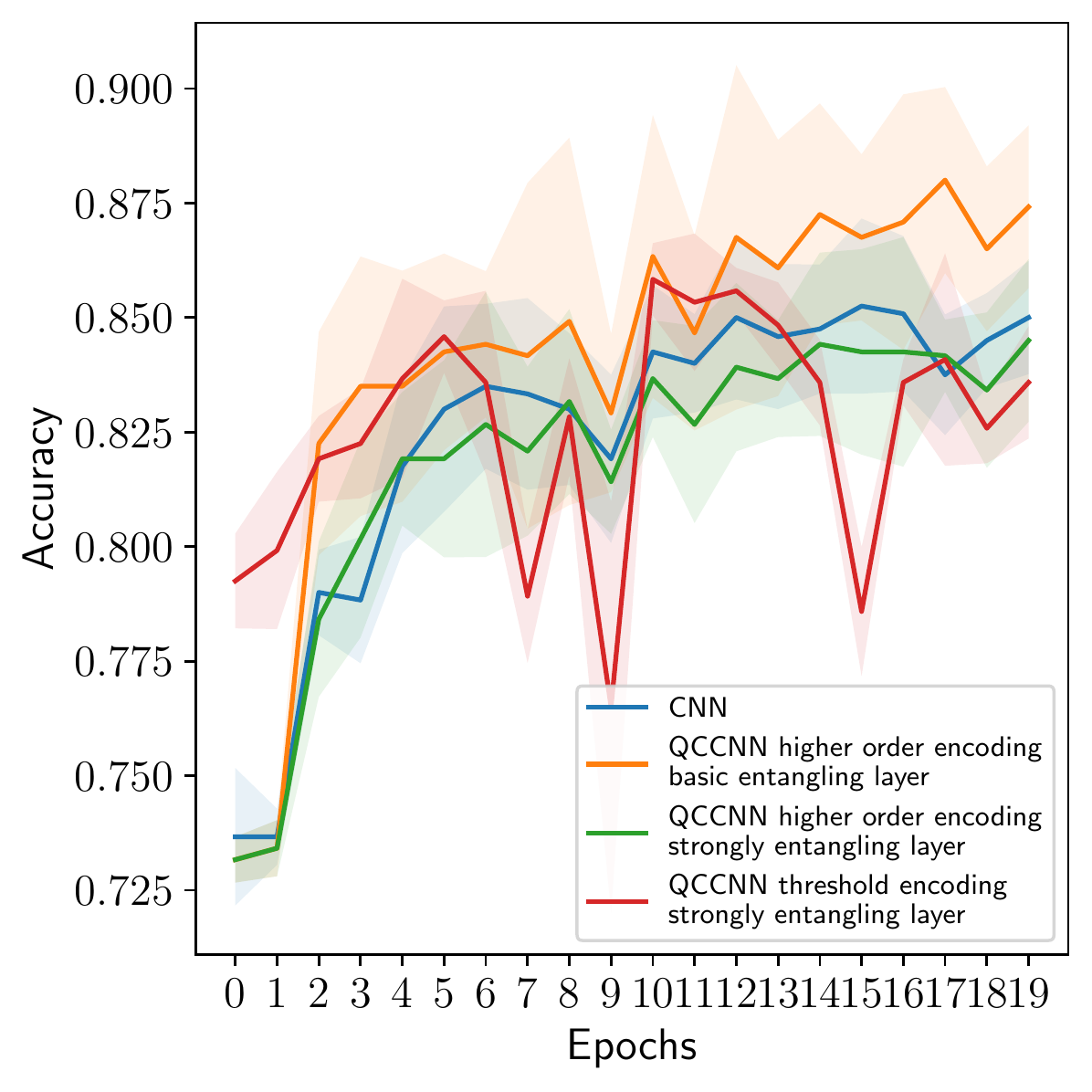}
    \caption{Validation accuracy}
\end{subfigure}
\qquad
\begin{subfigure}{0.4\textwidth}
    \centering
    \includegraphics[width=\textwidth]{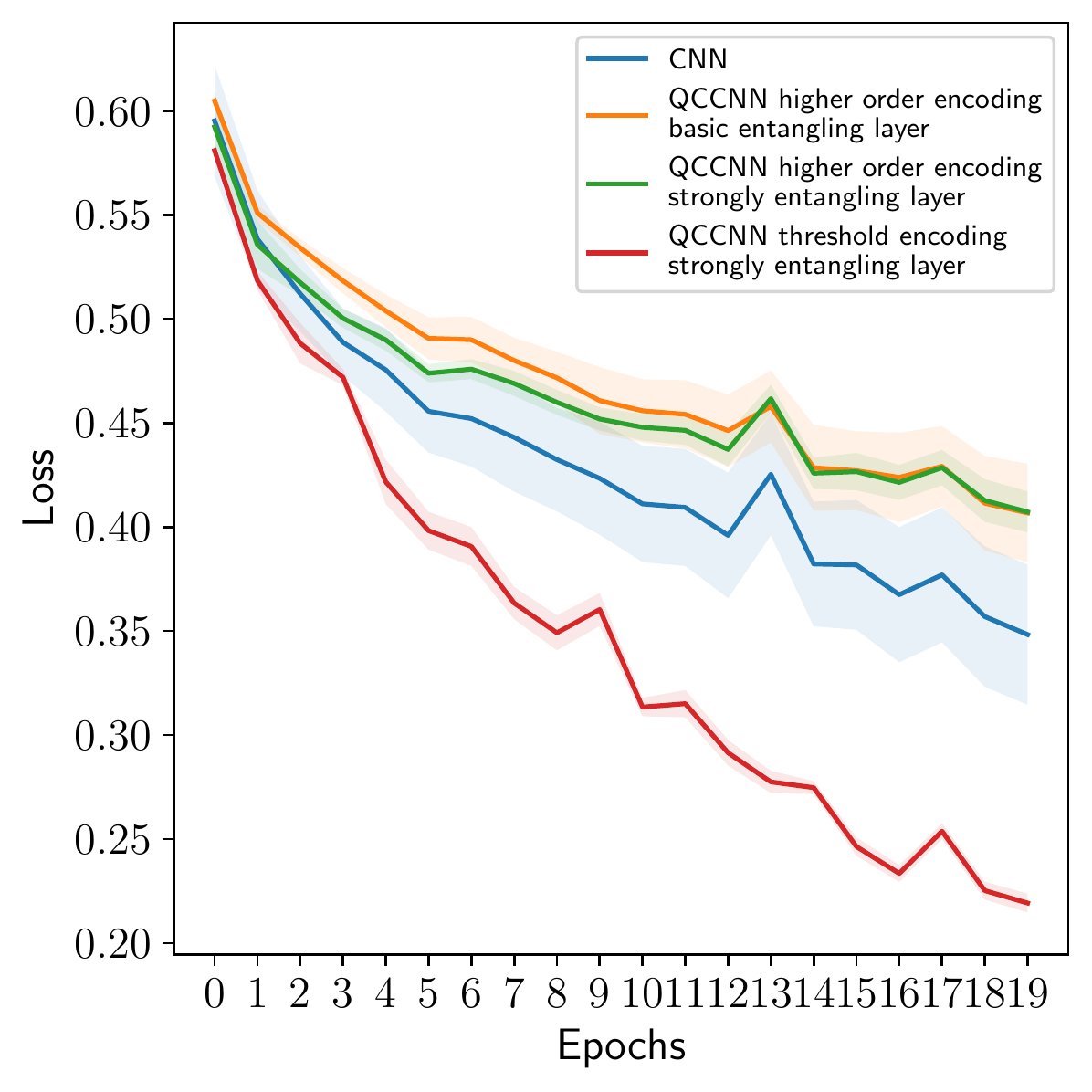}
    \caption{Training loss}
\end{subfigure}
\qquad
\begin{subfigure}{0.4\textwidth}
    \centering
    \includegraphics[width=\textwidth]{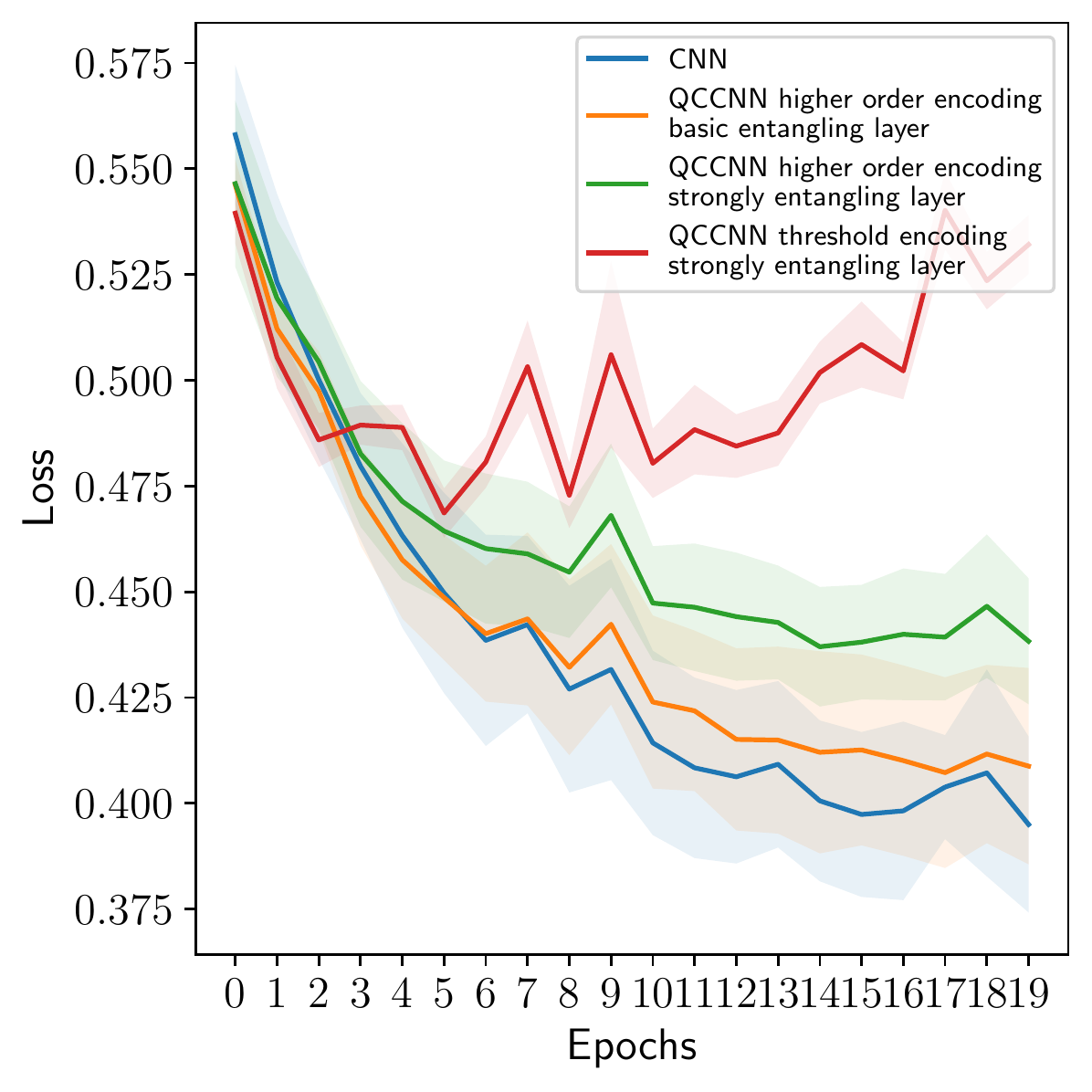}
    \caption{Validation loss}
\end{subfigure}

\caption{Hybrid model performance in terms of training and validation accuracy and cross-entropy loss compared to the performance of a classical CNN on the BreastMNIST dataset. The classical CNN (in blue) is compared to the hybrid QCCNNs with a basic entangling layer and with higher order encoding (in orange), a strongly entangling layer and with higher order encoding (in green), or a strongly entangling layer and with threshold encoding (in red).}
\label{fig:BreastMNIST}
\end{figure*}

\begin{figure*}[h!]
\centering

\begin{subfigure}{0.4\textwidth}
    \centering
    \includegraphics[width=\textwidth]{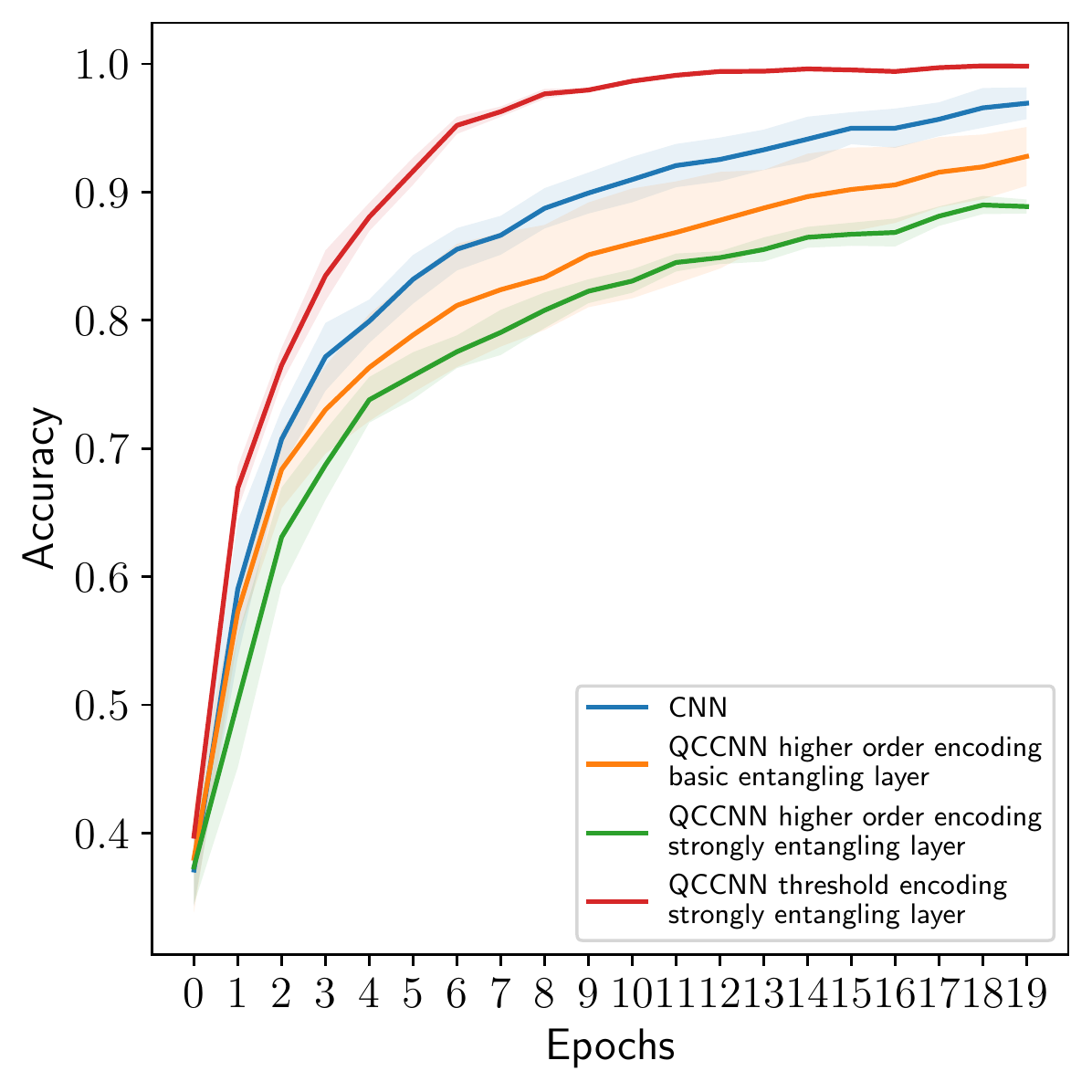}
    \caption{Training accuracy}
\end{subfigure}
\qquad
\begin{subfigure}{0.4\textwidth}
    \centering
    \includegraphics[width=\textwidth]{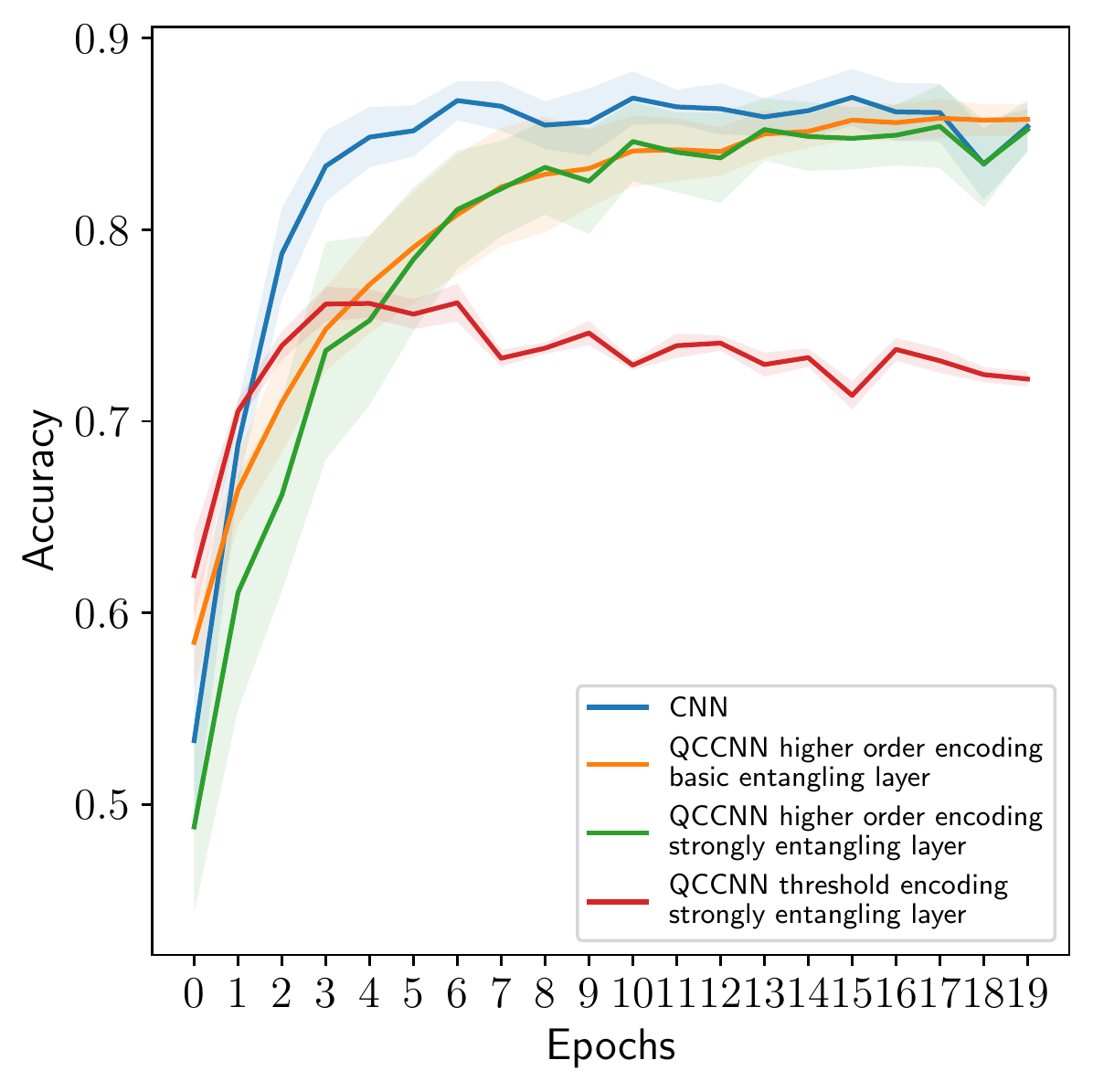}
    \caption{Validation accuracy}
\end{subfigure}
\qquad
\begin{subfigure}{0.4\textwidth}
    \centering
    \includegraphics[width=\textwidth]{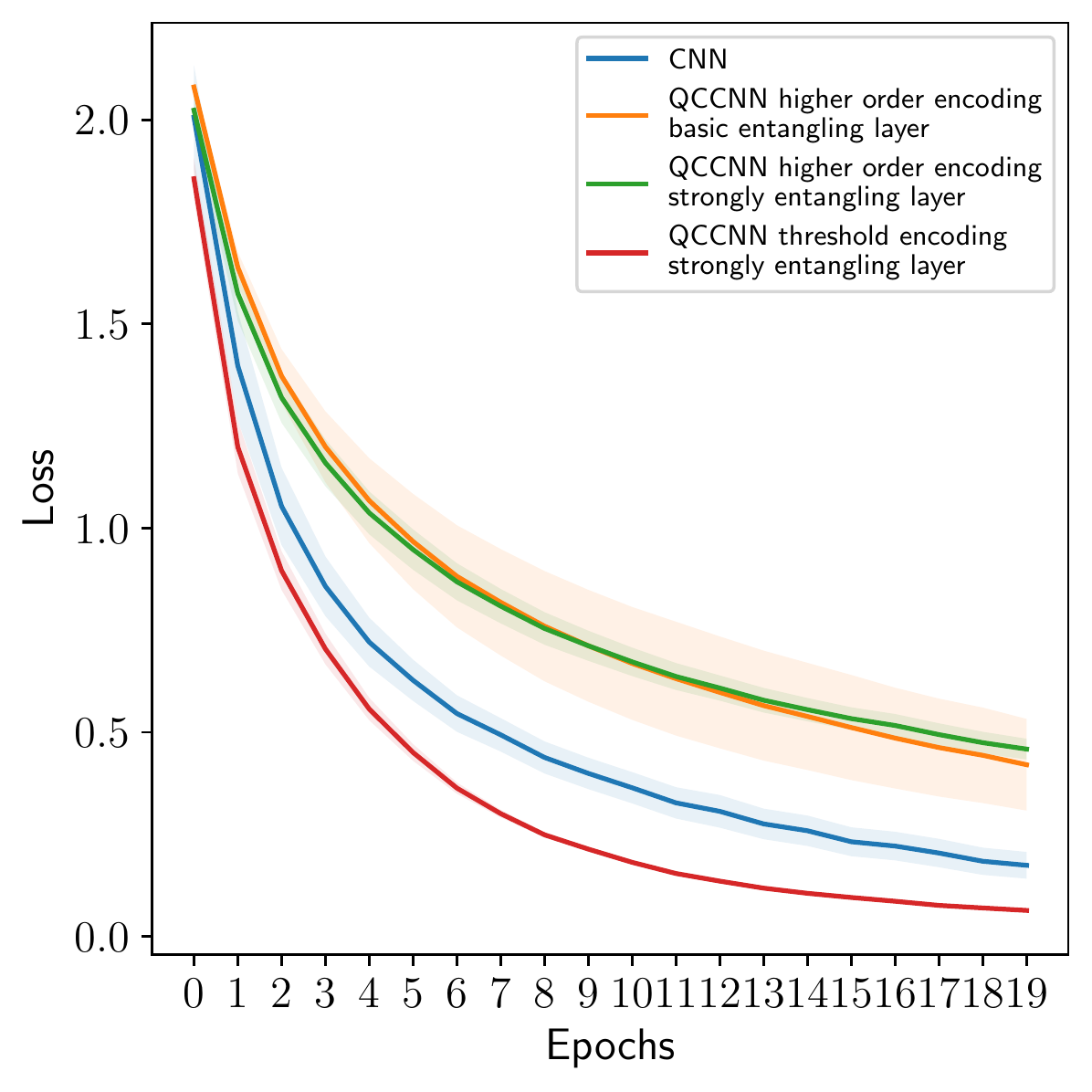}
    \caption{Training loss}
\end{subfigure}
\qquad
\begin{subfigure}{0.4\textwidth}
    \centering
    \includegraphics[width=\textwidth]{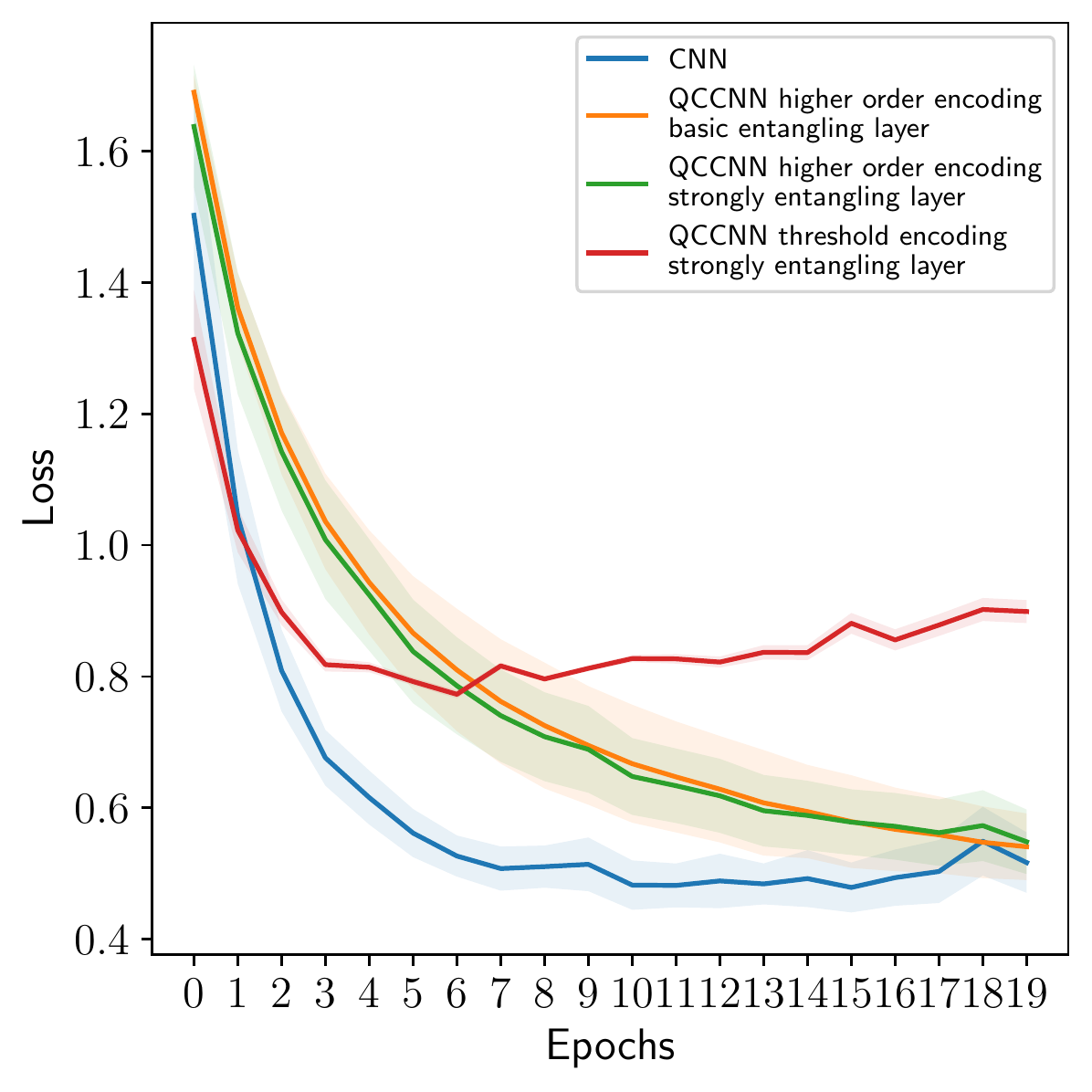}
    \caption{Validation loss}
\end{subfigure}

\caption{Hybrid model performance in terms of training and validation accuracy and cross-entropy loss compared to the performance of a classical CNN on the OrganAMNIST dataset. The classical CNN (in blue) is compared to three hybrid QCCNNs: with a basic entangling layer and with higher order encoding (in orange), a strongly  entangling layer and with higher order encoding (in green), or a strongly entangling layer and with threshold encoding (in red).}
\label{fig:OrganAMNIST}
\end{figure*}

%\twocolumngrid

For both 2D datasets the hybrid QCCNNs with higher order encoding perform comparably well with respect to their classical counterparts, as demonstrated by similar final validation accuracies in \figurename~\ref{fig:BreastMNIST} and \ref{fig:OrganAMNIST}. 

For the BreastMNIST dataset, while the classical CNN achieves a slightly lower mean loss than the QCCNNs with higher order encoding on the training set, the gap is closed on the validation set, where a similar loss is obtained with the classical network and the QCCNN with the basic entangling layer (\figurename~\ref{fig:BreastMNIST}). This indicates that the hybrid QCCNN with the higher order encoding and the basic entangling layer is able to generalize well on this small-scale dataset. Additionally, similar validation loss values do not necessarily lead to similar validation accuracies, as seen with the higher mean validation accuracy of the network with a basic entangling layer compared to the classical network. This shows that, on average, while the distances between the true labels and the predicted values by the model are similar, the QCCNN with the basic entangling layer made a smaller number of errors on the labels.

The QCCNNs with the higher order encoding trained on the OrganAMNIST dataset achieve a higher loss and a lower accuracy on average over the epochs compared to the classical CNN, but the validation metrics eventually converge to similar values, as seen in \figurename~\ref{fig:OrganAMNIST}. 

For both datasets, the QCCNN with threshold encoding largely overfits on the training data and poorly generalizes on the validation dataset. The highly unstable validation accuracy on the BreastMNIST dataset is most likely due to the small validation set. This unsatisfactory performance is expected on more complex data composed of shades of gray, as opposed to datasets such as the MNIST handwritten digits dataset~\cite{MNIST,FOKUS:2021}, since the threshold encoding into solely two states cannot represent the complexity of the features properly. 

Overall, when looking at the total number of trainable parameters, we reach a similar performance using less parameters, with 20 parameters in the classical convolutional layer, compared to only 4 parameters in the convolutional layer with the basic entanglement. This is a promising observation for future research. With its 12 trainable parameters, the QCCNN with a strongly entangling layer surprisingly shows slightly worse performance compared to the QCCNN with a basic entangling layer on both 2D datasets. This effect will also be investigated in  more depth in future work.

We additionally note that for the BreastMNIST dataset, the high variances of the observed metric values for the classical CNN and the QCCNNs show the strong dependency of the networks to the weights initialization. This effect probably arises from the small-scale dataset.
The large overlapping of the variance curves also highlights that the classical and hybrid networks' performances are indeed close. Regarding the OrganAMNIST dataset, with more data, we observe that the variance of the metrics is of the same order for the QCCNNs (except for the threshold encoding) compared to the previous dataset, but is considerably reduced on the classical CNN.  
\newline

\subsubsection{3D dataset}

\begin{figure*}[h!]
\centering

\begin{subfigure}{0.4\textwidth}
    \centering
    \includegraphics[width=\textwidth]{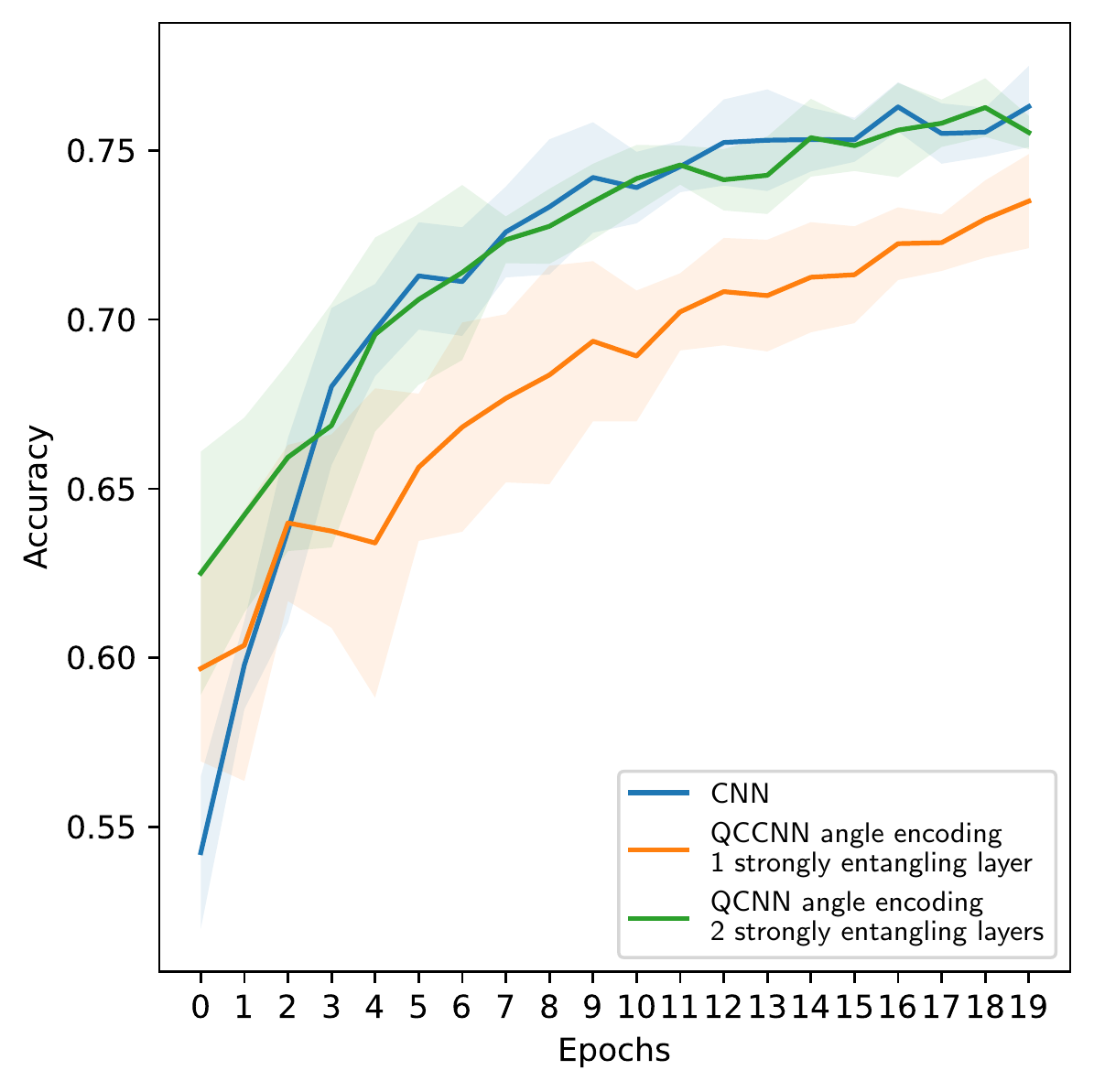}
    \caption{Training accuracy}
\end{subfigure}
\qquad
\begin{subfigure}{0.4\textwidth}
    \centering
    \includegraphics[width=\textwidth]{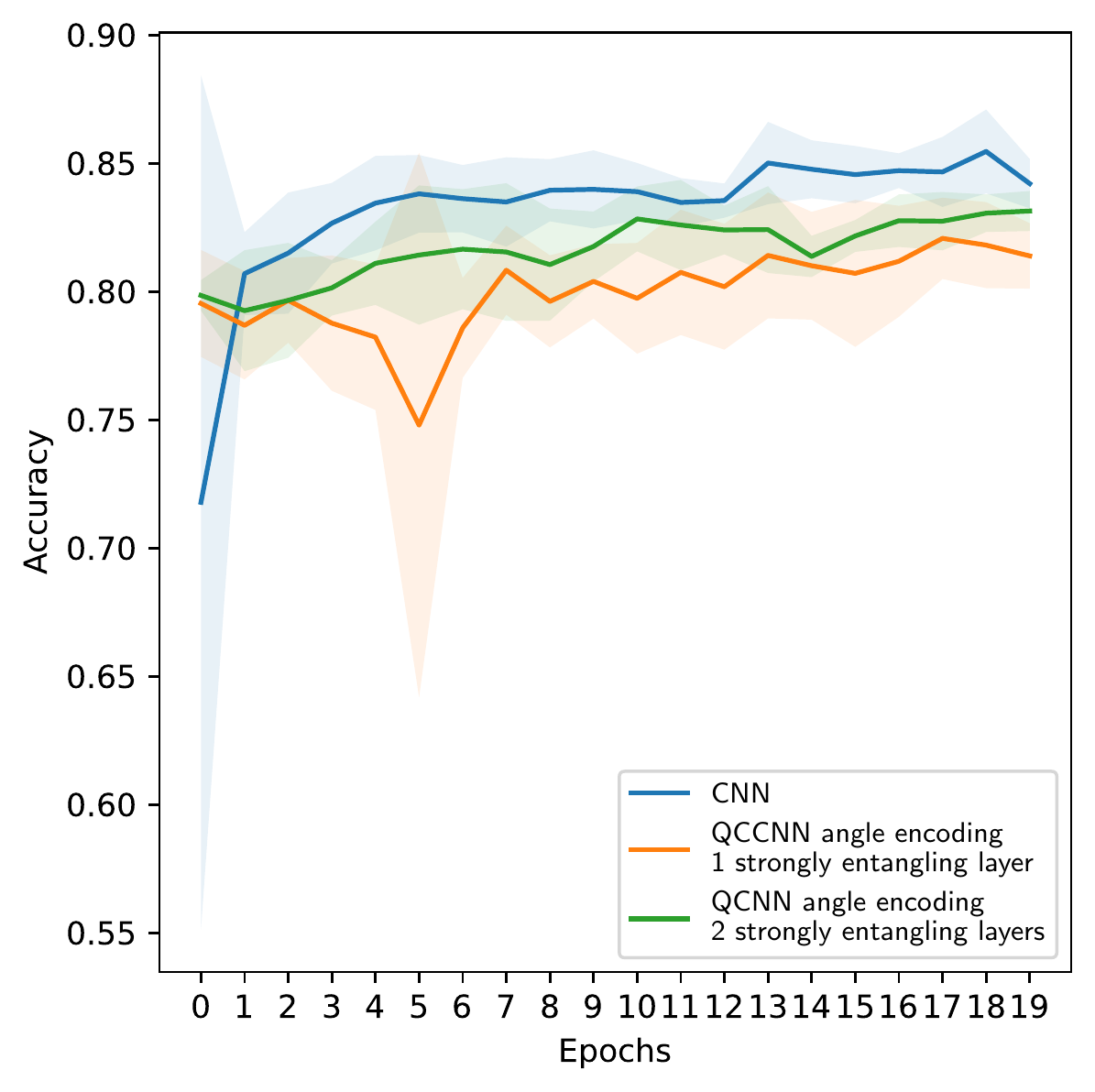}
    \caption{Validation accuracy}
\end{subfigure}
\qquad
\begin{subfigure}{0.4\textwidth}
    \centering
    \includegraphics[width=\textwidth]{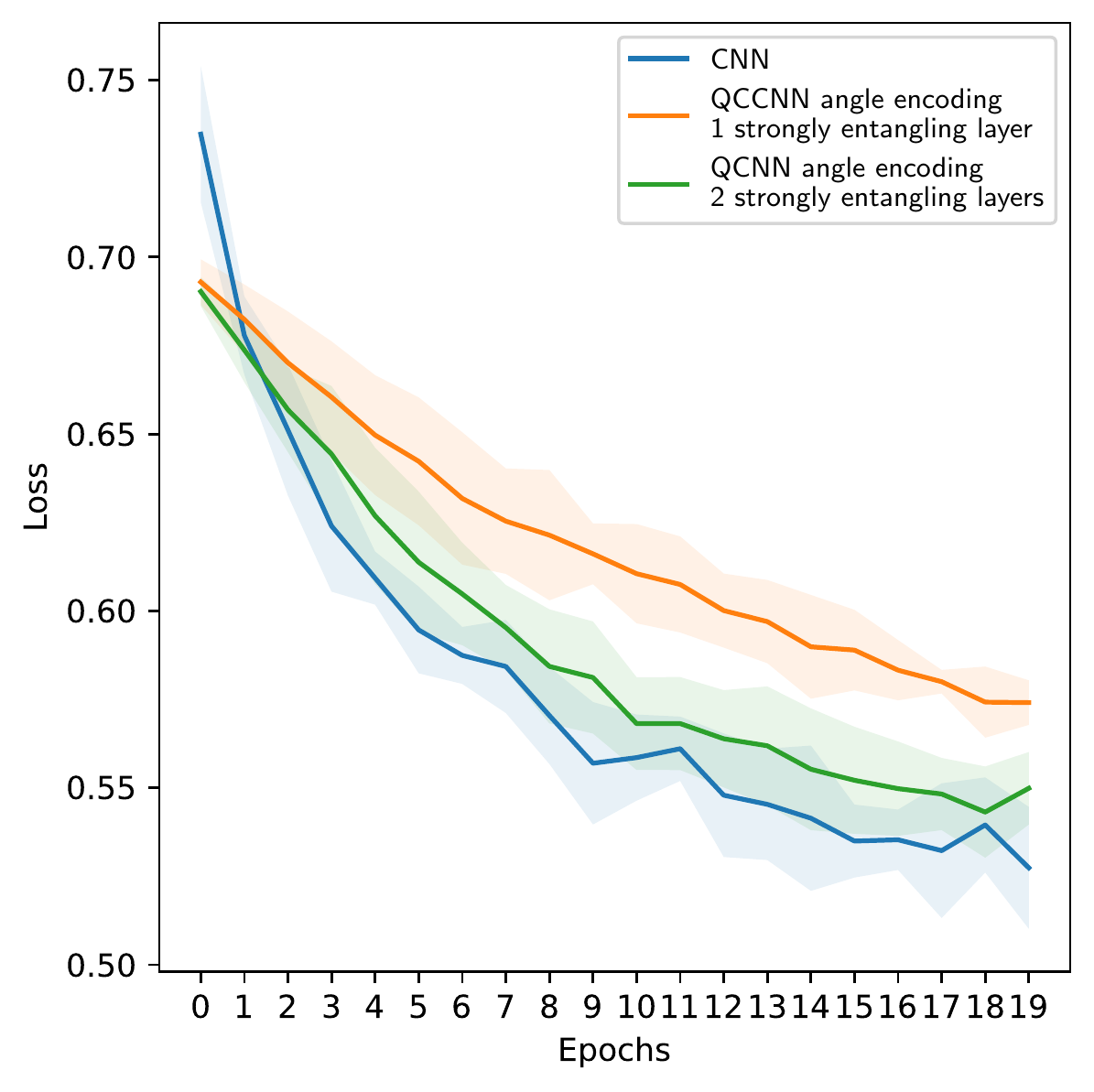}
    \caption{Training loss}
\end{subfigure}
\qquad
\begin{subfigure}{0.4\textwidth}
    \centering
    \includegraphics[width=\textwidth]{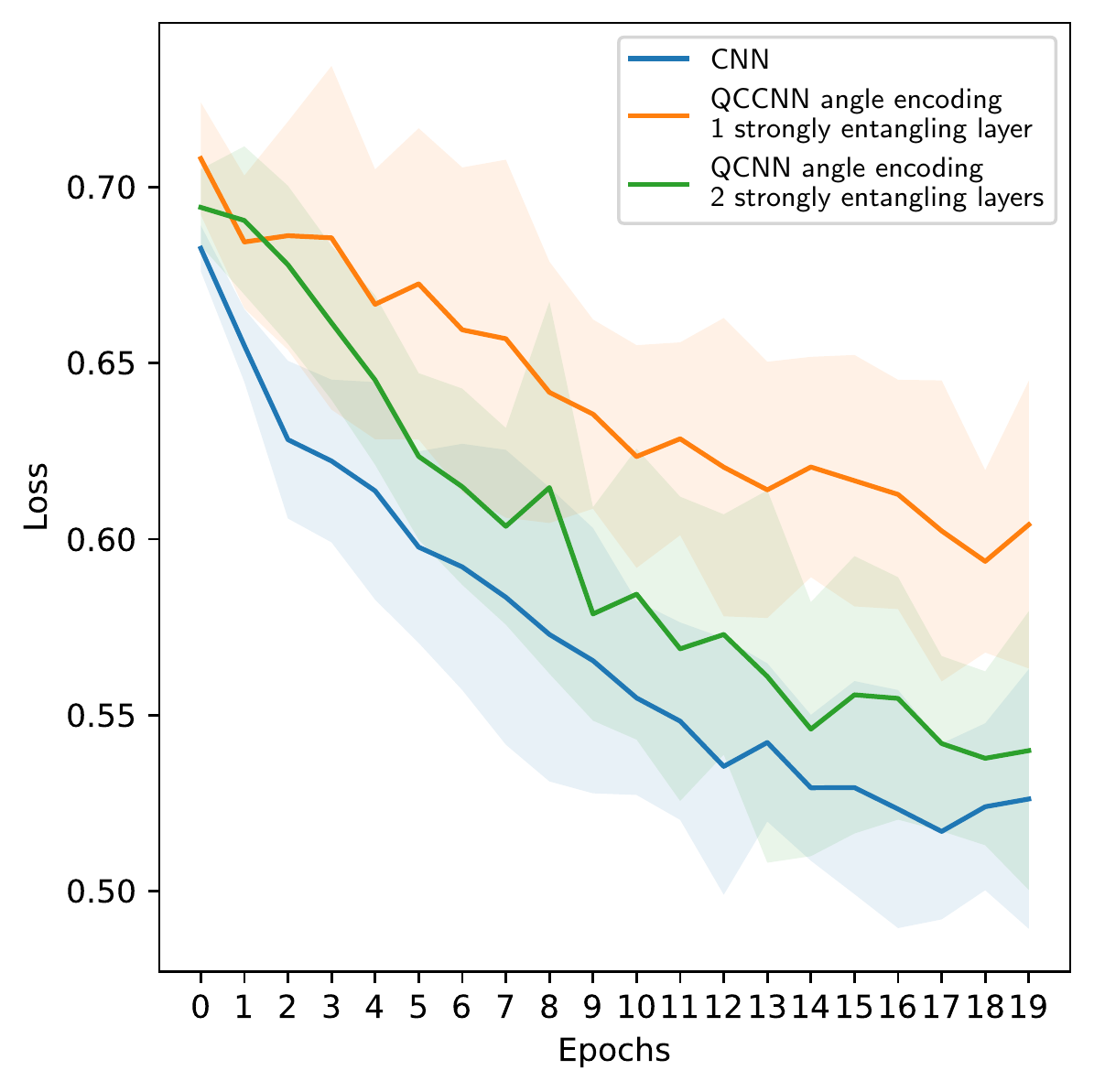}
    \caption{Validation loss}
\end{subfigure}

\caption{Comparison of the performance of the hybrid QCCNN model with the one of the classical CNN in terms of training and validation accuracy and cross-entropy loss on the 3D lung-nodule dataset. The classical CNN (in blue) is compared with QCCNNs using angle encoding and a strongly entangling layer (in orange) or with two strongly entangling layers (in green).}
\label{fig:3DCT}
\end{figure*}

The training and validation curves for the lung-nodule dataset are shown in \figurename~\ref{fig:3DCT}. After the third epoch, very similar training accuracy values are obtained for the classical CNN and the QCCNN with two strongly entangling layers. For the training loss as well as for the validation loss and accuracy, the best performance is achieved by the classical CNN. However, considering the relatively broad variance bands, the classical CNN and the QCCNN with two strongly entangling layers lead to a similar performance in general. Compared to them, the QCCNN with only one strongly entangling layer performs clearly worse during the training. For the validation dataset, a similar accuracy as for the QCCNN with two strongly entangling layers is obtained. However, the resulting validation loss is much larger than for the other two setups.

\section{Conclusion}

In conclusion, we obtain comparably good training results with QCCNNs as with classical CNNs on both 2D and 3D radiological image classification tasks.

With the 2D ultrasound images of the breast and the 2D abdominal CT images, the QCCNN using an encoding feature map with higher order encoding and a variational circuit consisting of a basic entangling layer presents a similar performance as the classical CNN, although a high uncertainty remains, represented by large training and validation variance bands. An alternative QCCNN with a variational map using a strongly entangling layer with three times more trainable parameters than the basic entangling layer, achieves worse performance. In future research, the quantum circuit design will systematically be studied to understand which set of gates leads to a better performance.
A clear advantage in training and generalization abilities of QCCNNs over classical CNNs in the presence of limited data as promised in~\cite{Caro:2021mgf} was not yet observed within this paper. Our results are nevertheless promising, as the number of parameters in the quantum convolutional layer is lower than in the classical convolutional layer. In future work, we will investigate the dependence of the learning ability on the parameter number and on the number of quantum convolutional filters used.

For the 3D lung-nodule dataset, we showed that with a quantum convolutional layer with angle encoding and two strongly entangling layers we can achieve a similar performance as with a classical CNN. In future research it should be studied whether a more suitable QCCNN can be constructed to further increase the performance, e.g. through a different circuit design. Especially, it needs to be investigated which encoding strategy is the most suitable one for 3D data, in particular considering high-resolution medical data required in real applications. Further research is also required on the encoding strategy for QCCNNs, in which the quantum convolutional layer is placed after one or more classical convolutions.

\section*{Acknowledgment}
The authors acknowledge the National Cancer Institute and the Foundation for the National Institutes of Health, and their critical role in the creation of the free publicly available LIDC/IDRI Database used in this study. A. Matic and M. Monnet contributed equally to the work. The authors thank Alona Sakhnenko for providing the circuit sketches in \figurename~\ref{fig:circuit_2D} and \ref{fig:circuit_3D}.

\end{document}